\documentclass[12pt]{article}

\usepackage{geometry}
\usepackage{graphicx} 
\usepackage{amsmath, amssymb}
\usepackage{natbib}
\usepackage{booktabs}
\usepackage{sectsty} 
\usepackage{helvet}
\usepackage{enumerate}
\usepackage{url}
\usepackage{xcolor}
\usepackage{subfigure}

\newcommand{\dd}[1]{\mathrm{d}{#1}}
\renewcommand{\vec}[1]{\pmb{#1}}
\newcommand{\X}{\vec{x}}
\newcommand{\Y}{\vec{y}}
\newcommand{\Z}{\vec{z}}
\newcommand{\R}{\vec{R}}
\newcommand{\T}{\vec{t}}
\newcommand{\KC}{\kappa}
\newcommand{\upperbound}{u}
\newcommand{\nsigma}{3}
\newcommand{\err}{\mathrm{error}}

\DeclareMathOperator*{\argmin}{argmin}

\usepackage[helvet]{sfmath}

\title{Matching biomolecular structures by registration of point clouds}
\author{Michael Habeck\,$^{\text{\sf 1,2}*}$, 
  Andreas Kr\"opelin\,$^{\text{\sf 1}}$, and
  Nima Vakili\,$^{\text{\sf 1,2}}$}
\newcommand{\myaffiliation}{%
  \noindent $^{\text{\sf 1}}$Microscopic Image Analysis Group, Jena University Hospital, 07743 Jena, Germany\\
  $^{\text{\sf 2}}$Max Planck Institute for Multidisciplinary Sciences, 37077 G\"ottingen, Germany\\\\
}
\newcommand{\myemail}{*E-Mail: michael.habeck@uni-jena.de}

\begin{document}

\maketitle
\myaffiliation
\myemail

\abstract{
\noindent\textbf{Motivation:} Assessing the match between two biomolecular structures is at the heart of structural analyses such as superposition, alignment and docking. These tasks are typically solved with specialized structure-matching techniques implemented in software for protein structural alignment, rigid-body docking, or rigid fitting into cryo-EM maps. \\ 
\textbf{Results:} We present a unifying framework to compare biomolecular structures by applying ideas from computer vision. The structures are represented as three-dimensional point clouds and compared by quantifying their overlap. We use the kernel correlation to measure point cloud overlap, and discuss local and global optimization strategies for maximizing the kernel correlation over the space of rigid transformations. We derive a majorization-minimization procedure that can be used to register two point clouds without establishing a point-to-point correspondence. We demonstrate that the majorization-minimization algorithms outperform the commonly used Iterative Closest Point registration algorithm. Furthermore, we discuss and benchmark a a randomization strategy for globally optimizing the kernel correlation. We illustrate the approach on various 3D fitting problems such as the comparison of circularly permuted structures and rigid fitting of cryo-EM maps or bead models from small-angle scattering.
}
\clearpage

\section{Introduction}

Superposition and comparison of biomolecular structures are common tasks in structural biology. Structures are compared and aligned to predict the conformation and function of proteins, for example by homology modeling. To assess the conformational diversity of structures revealed by NMR or other structure determination methods, multiple conformations need to be superimposed in a meaningful fashion. Rigid docking approaches optimize the match between an experimental shape revealed by cryo-electron microscopy or solution scattering with a known structure.

Aside from manual superposition approaches requiring user intervention, the standard approach to superimpose and compare different conformations of the same bio\-molecule is to minimize their root-mean-square deviation (RMSD). To compute the RMSD, six rigid degrees of freedom are determined by a singular value decomposition \citep{Kabsch76}. This assumes a least-squares criterion for assessing the match between two structures. 

Robust variants of the RMSD take account of the fact that the RMSD will be less meaningful if the structures undergo large conformational changes \citep{Hirsch08,Mechelke10} or show varying degrees of structural heterogeneity \citep{Theobald06}. Variants of the standard least-squares superposition such as the weighted RMSD \citep{Damm06} have been proposed to tackle more challenging superposition tasks. 

Optimization of the (weighted) RMSD is feasible by extending the Kabsch algorithm, but restricted to cases where the correspondence between the positions in both structures is known. This requirement holds, for example, for NMR ensembles. In the more general case, the correspondence between positions in both structures is unknown. Both problems, 3D superposition and establishing a correspondence between equivalent positions, are intertwined and cannot be solved independently of each other. 

This situation occurs in protein structural alignment where we need to solve two problems: First, we need to establish the correspondence between evolutionarily related amino acids. Second, we must find the best superposition of corresponding three-dimensional coordinates such that related amino acids are close in space. However, the approach is restricted to situations where the correspondence is consistent with the sequential order of the amino acids (such that dynamic programming can be used to solve the alignment problem). But there are cases where standard structural alignment of evolutionarily related proteins does not apply any more such as, for example, circular permutation, 

When working with reconstructions from cryo-electron microscopy (cryo-EM), we might want to dock a high-resolution structure into a 3D density map \citep{Villa14}. Rigid docking could either use a voxel-based representation of both structures or particle models such as an atomic structure or a bead model. In the first case, the high-resolution structure needs to be converted to a density map. In the second case, the density map has to be converted to a (pseudo)-atomic structure. \citet{Kawabata08} introduced a decomposition of cryo-EM density maps into a mixture of anisotropic Gaussians that are characterized by a center position, a weight and a covariance matrix (representing an ellipsoidal shape). Omokage search \citep{suzuki2016omokage} uses this representation to rapidly compare the overall shape of two or more structures obtained with X-ray crystallography, cryo-EM or small-angle scattering (SAS).

Here, we solve all of the mentioned comparison and superposition tasks within a common framework. We use a particle-based representation of biomolecular structures including atomic structures, cryo-EM density maps, and bead models from SAS. To compare two biomolecular structures, we use the {\em kernel correlation} which has been introduced in computer vision to register point clouds. We discuss local and global strategies to optimize the kernel correlation over rigid transformations of one structure against the other structure. Finally, we illustrate our approach on various comparison and superposition tasks. 

\section{Methods}

\subsection{Representation of biomolecular structures by weighted point clouds}
We use weighted point clouds to represent biomolecular structures from different experimental sources. A weighted point cloud comprises a collection of $I$ points at positions $\X_i$ and with associated weights $p_i$. The positions can be stored in an $I\times 3$ matrix $\vec{X}$ whose rows are the 3D coordinates of the points, the weights form an $I$-dimensional vector $\vec{p}$. 

Atomic structures \citep{Berman00} can naturally be viewed as point clouds where alpha carbons often serve as representative atoms and weights could be constant or proportional to the mass or occupancy. In case of large structures such as multi-domain proteins or macromolecular complexes, coarse graining allows us to reduce the size of the point cloud such that a single point represents multiple atoms.

Cryo-EM structures are typically represented as gridded 3D volumes. Structural manipulations such as a rigid transformation necessitate the interpolation of voxels values, which is slow and prone to artifacts. We therefore prefer to also represent volumetric data as weighted point clouds \citep{vakili2021bayesian}. A powerful algorithm to obtain weighted point clouds from large atomic structures and cryo-EM maps is DP-means \citep{Kulis12}, a non-parametric version of the K-means algorithm. In DP-means, the radius of the bead is chosen by the user and the number of points is estimated automatically.

\subsection{Assessing the match between two point clouds by the kernel correlation}
Let us now compare two point clouds $(\vec{X}\!, \vec{q})$ and $(\vec{Y}\!, \vec{p})$ of size $I$ and $J$, respectively. Both clouds are typically represented in different frames of reference, and we need to rigidly transform one cloud, the {\em source}, against the other, the {\em target}, in order to compare them meaningfully. This is a common task in computer vison called {\em rigid registration}. A rigid transformation involves a $3\times 3$ rotation matrix $\R$ and translation vector $\T$. To find the best pose $(\R, \T)$, we must optimize a quantitative measure of how well two point clouds match. The {\em kernel correlation} (KC) $\KC$ \citep{Tsin04}
\begin{equation}\label{eq:KC}
  \KC(\R, \T)
  = \sum_{i=1}^I \sum_{j=1}^J q_i\, p_j\, \phi(\|\X_i - \R\Y_j - \T\|)
  = \vec{q}^T \pmb \Phi(\R, \T)\, \vec{p}
\end{equation}
is such a measure where $\phi$ is a suitable kernel, $\|\cdot\|$ the Euclidean norm and $\vec{\Phi}$ the $I\times J$ kernel matrix with elements $\phi_{ij} = \phi(\|\X_i - \R \Y_j - \T\|)$. Here, the weighted point cloud $(\vec{X}, \vec{q})$ is the fixed target, whereas $(\vec{Y}, \vec{p})$ is the movable source. KC is invariant under permutation of the point indices, and does not require a correspondence between the points in both clouds. This is convenient because optimization of KC will align two structures without knowing or assuming a point-to-point correspondence. 

Throughout this paper, we use the Gaussian radial basis function (RBF) kernel
\begin{equation}\label{eq:gaussian}
  \phi_\sigma(r)
  = \frac{1}{(2\pi\sigma^2)^{3/2}} \exp\left( - \tfrac{1}{2\sigma^2} r^2 \right)
\end{equation}
where the bandwidth $\sigma > 0$ determines how tolerant KC is against mismatches. The correlation of two Gaussian kernels is
\begin{equation}\label{eq:self-reproducing}
  \left\langle
  \phi_{\sigma_1}(\|\vec{\cdot} - \vec{\mu}_1\|),
  \phi_{\sigma_2}(\|\vec{\cdot} - \vec{\mu}_2\|)
  \right\rangle
  = \phi_\sigma(\|\vec{\mu}_1 - \vec{\mu}_2\|)
\end{equation}
where $\langle q, p \rangle = \int q(\X) p(\X) \dd\X$ denotes the functional inner product; the variances satisfy $\sigma^2 = \sigma_1^2 + \sigma_2^2$. Due to the self-reproducing property of the Gaussian kernel (Eq. \ref{eq:self-reproducing}), KC can be viewed as the inner product 
\begin{equation}
  \KC(\R, \T)
  = \left\langle
  q_{\sigma_1}, p_{\sigma_2}\bigl(\R^T (\,\,\vec{\cdot}\, - \T) \bigr)
  \right\rangle
\end{equation}
of kernel density estimates (KDEs)
\begin{equation}\label{eq:KDE}
q_\sigma(\X) = \sum_{i=1}^I q_i\, \phi_{\sigma}(\|\X - \X_i\|), \quad
p_\sigma(\X) = \sum_{j=1}^J p_j\, \phi_{\sigma}(\|\X - \Y_j\|)
\end{equation}
whose variances satisfy $\sigma^2 = \sigma_1^2 + \sigma_2^2$. By maximizing KC (Eq. \ref{eq:KC}), we minimize the squared distance between the KDEs (Eq. \ref{eq:KDE}). 

The Kpax algorithm by \citet{Ritchie16} uses a related match criterion for protein structure alignment. The major difference is that Kpax's objective function is not the total sum over all elements of the kernel matrix as in Eq. (\ref{eq:KC}), but reduced to the aligned point pairs. For general point cloud comparison, establishing an alignment is no longer suitable (think of circularly permuted protein structures or cryo-EM maps). 

An important parameter is the bandwidth $\sigma$. The smaller $\sigma$ the rougher and more difficult to optimize will be KC. On the other hand, large $\sigma$ will result in ambiguous registrations. Larger values of $\sigma$ are more suitable for global registration, whereas a small $\sigma$ allows us to find locally similar subsets of points in both clouds. In principle, $\sigma$ is a free parameter that can be chosen by the user or by methods used in kernel density estimation. In our applications, we exploit the fact that we are comparing biomolecular structures and therefore can use our domain knowledge to fix reasonable $\sigma$ values. For example, when working with cryo-EM maps, the resolution of the map gives an estimate for the appropriate $\sigma$. For the  comparison of alpha carbon clouds derived from atomic structures, we typically use $\sigma=5 \,\text\AA \approx \sqrt{2}\times 3.5 \,\text\AA$ where 3.5 \AA{} is roughly the average distance between alpha carbons. \citet{Ritchie16} uses a smaller bandwidth, $\sigma = \sqrt{2}\times 1.4 \,\text\AA$ for aligning protein structures. 

\subsection{Fast evaluation of the kernel correlation}\label{sec:grid}
The evaluation of KC (Eq. \ref{eq:KC}) scales with $IJ$, which impedes systematic searches for the globally optimal pose. However, restricting the Gaussian kernel to a finite support can result in significant speed-ups. We typically use a $3\sigma$ cutoff: $\phi_\sigma(r) = 0$ for $r > 3 \sigma$. As a consequence, the double-sum in the evaluation of KC (Eq. \ref{eq:KC}) can be restricted to a sum over contributions from points in a finite neighborhood.  Efficient data structures for spatial queries such as $k$-d trees, ball trees or neighbor lists can be used to rapidly determine nearest neighbors or contacts below the cutoff distance. This allows us to compute accurate approximations of KC in a fraction of time. 

Grid-based techniques can produce further speed-ups. KC can be seen as the inner product of a blurry target density $q_\sigma$ (Eq. \ref{eq:KDE}) and a sharp source density $p_0(\X) = \sum_{j=1}^J p_j \, \delta(\|\X - \Y_j\|)$ where $\delta(\cdot)$ denotes the Dirac delta function. Therefore, we can approximate KC by a vector inner product over a grid $\mathcal G$ of 3D positions:
\begin{equation}\label{eq:grid-KC}
\KC(\R, \T)
= \left\langle q_\sigma, p_0\bigl(\R^T (\,\,\vec{\cdot}\, - \T) \bigr) \right\rangle
\approx \sum_{\vec{ x} \in \mathcal G} q_\sigma(\X) \, p_0\bigl(\R^T(\X - \T)\bigr)\, .
\end{equation}
We use a cubic rectilinear grid $\mathcal G$ to discretize the inner product. The vector  $\{q_\sigma(\X)\}_{\X \in \mathcal G}$ needs to be computed only once in a search over multiple rotations and translations. The vector representing the source in pose $(\R, \T)$ is obtained by subtracting the grid origin from the transformed points $\R\Y_j + \T$, followed by a division of the shifted points by the grid spacing and subsequent rounding. This results in a one to three orders of magnitude faster computation of the kernel correlation. A comparison of the computation times achieved with these implementations of the kernel correlation is shown in Supplementary Figure \ref{fig:times}. 

\subsection{Local optimization of the kernel correlation by iterative\\ majorization-minimization (MM)}\label{sec:MM}
Minimization of $- \log \KC$ is equivalent to maximizing KC and produces an optimal rigid registration of two point clouds. This is a non-convex optimization problem with many local minima corresponding to partial matches of both point clouds. To optimize the scoring function $-\log\KC$, we construct an upper bound that can be minimized in closed form:
\begin{eqnarray*}
  - \log \KC(\R, \T)
  &=& - \log \sum_{ij} q_i\, p_j\, \phi_\sigma(\|\X_i - \R\Y_j - \T\|) \\
  &\le& - \sum_{ij} q_i\, p_j\, w_{ij} \log \frac{\phi_\sigma(\|\X_i - \R\Y_j - \T\|)}{w_{ij}} \\
  &=& \frac{1}{2\sigma^2} \sum_{ij} q_i p_j w_{ij} \|\X_i - \R \Y_j - \T\|^2 +
  \mathrm{const}
\end{eqnarray*}
where we used Jensen's inequality. The constant, $\sum_{ij} q_i p_j w_{ij} \log w_{ij}$, does not depend on the parameters of the rigid transformation. The inequality is valid for all weights $w_{ij}$ satisfying $w_{ij} \ge 0$ and $\sum_{ij} q_i p_j w_{ij} = 1$. For weights proportional to the kernel matrix, $w_{ij} \propto \phi_\sigma(\|\X_i - \R \Y_j - \T\|)$, the upper bound touches $-\log\KC$ at $(\R, \T)$, and the inequality becomes an identity.

This suggests a majorization-minimization (MM) strategy \citep{Hunter04} to optimize KC by cycling between updates of $w_{ij}$ followed by minimization of the upper bound
\begin{equation}\label{eq:upper}
  \upperbound(\R, \T) =
  \frac{1}{2\sigma^2} \sum_{ij} q_i p_j w_{ij} \|\X_i - \R \Y_j - \T\|^2\, .
\end{equation}
The solution of $\argmin\, \upperbound(\R, \T)$ is available in closed form (see supplementary information). The optimal translation is
\begin{equation}\label{eq:translation}
  \vec{\hat\T} = \vec{\overline x} - \vec{\hat\R} \,\vec{\overline y}
\end{equation}
with centers of mass $\overline{\X} = \sum_{ij} q_i p_j w_{ij} \X_i$ and $\vec {\overline y} = \sum_{ij} q_i p_j w_{ij} \Y_j$. The optimal rotation can be computed by solving the matrix nearness problem
\begin{equation}\label{eq:rotation}
  \vec{\hat\R} = \argmin_{\R\, \in\, SO(3)} \|\R - \vec{S}\|^2_F\quad
\end{equation}
where $\vec{S} = \sum_{ij} q_i p_j w^{(n)}_{ij} (\X_i - \overline{\X}) (\Y_j - \overline{\Y})^T$ and $\|\cdot\|_F$ is the Frobenius norm. Minimization problem (\ref{eq:rotation}) can be solved by singular value decomposition of the $3\times 3$ matrix $\vec{S}$ \citep{Higham89}.

The following iterative MM procedure minimizes the negative logarithm of the kernel correlation:
\begin{itemize}
\item Initialization: Generate a random pose $(\R, \T)$ (alternatively, we can try to find a good initial pose by some heuristic). 
\item Iterate until convergence (e.g. when changes in $-\log\KC$ are no longer significant) or until a maximum number of iterations has been reached: 
  \begin{enumerate}
  \item Evaluate the kernel matrix $\phi_{ij} = \phi_\sigma(\|\X_i - \R \Y_j - \T\|)$ at the current pose $(\R, \T)$ and compute the normalized weights $w_{ij} = \phi_{ij} / \sum_{i'j'} q_{i'} p_{j'} \phi_{i'j'} = \phi_{ij} / \KC(\R, \T)$.
  \item Minimize the upper bound $\upperbound(\R, \T)$ by calculating the optimal rotation $\vec{\hat\R}$ and translation $\vec{\hat\T}$ according to equations (\ref{eq:rotation}) and (\ref{eq:translation}).
  \end{enumerate}
\end{itemize}
Supplementary Figure \ref{fig:upper_bound} illustrates the MM iterations for a specific example. It is also possible to implement the MM updates when using a grid approximation of the kernel correlation (Eq. \ref{eq:grid-KC}); see the supplementary information for details. 

\begin{figure}%[hbt!]
  \centering
  \includegraphics[width=\textwidth]{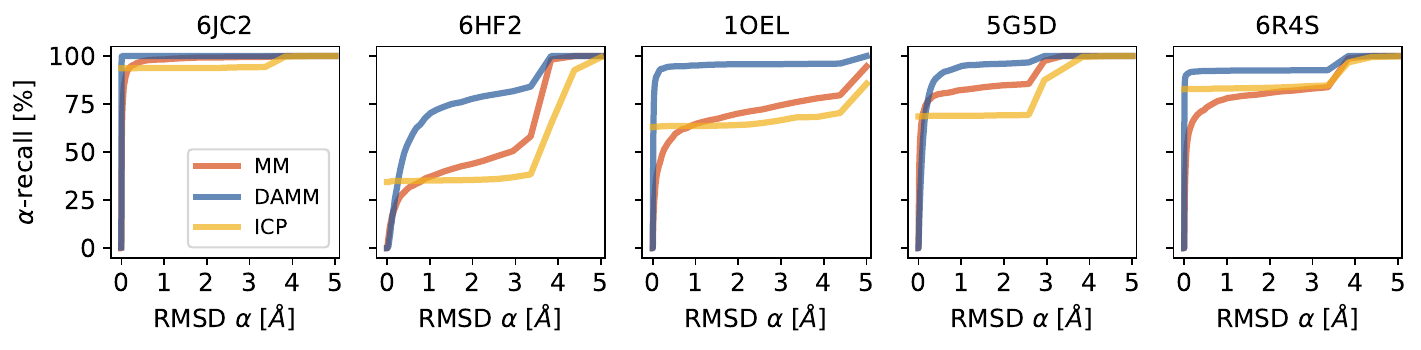}
  \caption{
    Testing local optimization strategies on various self-matching problems. A PDB structure (indicated in panel titles) is fitted against a permuted and randomly transformed version of itself. The performance is evaluated in terms of the $\alpha$-recall, which is the percentage of tests for which a given method achieves an RMSD below $\alpha$.}
  \label{fig:local}
\end{figure}
\subsection{Deterministic annealing}\label{sec:annealing}
Choosing the kernel width $\sigma$ should not be seen as a burden, but as a means to incorporate prior knowledge and control the shape of the objective function $-\log\KC$. One of the most widely used methods for rigid registration is the Iterative Closest Point (ICP) algorithm  \citep{Besl92,Chen92}. Like our MM approach, ICP iterates over two elementary steps: First, ICP establishes a correspondence between points in $\vec{X}$ and $\vec{Y}$ by matching pairs whose distance $\|\X_i - \R \Y_j - \T\|$ is minimal. Second, the least-squares fitting problem is solved for all pairs of corresponding points. However, ICP lacks a bandwidth parameter, its only algorithmic parameter is the number of iterations. 

In the KC approach, we can use the bandwidth $\sigma$ to gradually change the objective function. Because MM is only a local search strategy, it will strongly depend on the initial pose, which also holds for ICP. To avoid getting trapped in the nearest pose, we propose a simple modification reminiscent of deterministic annealing \citep{Rose90}. During iterative MM, we decrease the kernel bandwidth gradually until we reach the desired $\sigma$ value. The bandwidth is analogous to a temperature: Large $\sigma$ values (high temperatures) result in a flat cost function with shallow minima, annealing (reduction of $\sigma$) makes the cost function rougher, but also more selective. In our tests, we chose a simple linear annealing schedule starting at a large $\sigma_{\max}$ (typically 15 \AA{} or more generally $5\sigma$) and decrease the kernel bandwidth by a constant increment in each iteration. Obviously there are more options for the initial bandwidth and the progression of annealing. We found that the simple linear temperature schedule is sufficient for the registration problems considered in this article. We will use the abbreviation DAMM to denote the combination of iterative majorization-minimization and deterministic annealing. 

\section{Results}
We first report tests on local and global optimization strategies and then apply point cloud registration techniques to various structure comparison and fitting tasks. 

\subsection{Performance of local registration methods on a self-\\ matching benchmark}
The following structure matching task sheds some light on the strengths and shortcomings of the local optimization techniques detailed in Methods: A PDB structure is converted to a point cloud of alpha carbon positions with weight one. This point cloud serves as target against which a modified version of the same point cloud is matched. To generate the source point cloud, we apply a random permutation and rigid transformation to the target. Since the kernel correlation is invariant under permutation of the point indices, the optimal rigid registration will produce the same kernel correlation that is achieved by matching the target against itself: $\sum_{ij} \phi_\sigma(\|\X_i - \X_j\|)$. We set $\sigma=5$ \AA{} and run self-matching tests on structures with PDB codes 6JC2 (212), 6HF2 (325), 1OEL (524), 5G5D (160), 6R4S (382), where the numbers in brackets indicate the number of carbon alpha positions.

We test the local registration techniques, i.e. the MM algorithm (subsection \ref{sec:MM}) and its annealed version DAMM (subsection \ref{sec:annealing}), and compare them to ICP. For each structure, we generate 1000 self-matching problems by randomly shuffling the positions of the target and transforming them by a random rotation and translation. Ten random initial poses are generated from which each of the local optimization methods (MM, DAMM, and ICP) starts and runs for 50 iterations. The success of the registration method is assesed by computing the root mean square deviation (RMSD) between corresponding points in the target and transformed source. For a given pose $(\R, \T)$, the RMSD is defined as
\begin{equation}\label{eq:rmsd}
  \text{RMSD} = \sqrt{\frac{1}{I}
    \sum_{i=1}^I \min_{1 \le j \le J} \{\|\X_i - \R \Y_j - \T\|^2\} }
\end{equation}
For the self-matching tasks, the optimal RMSD is zero. The error defined in Eq. (\ref{eq:rmsd}) is the objective function of the ICP algorithm. 

\begin{figure}
  \centering
  \includegraphics[width=0.5\textwidth]{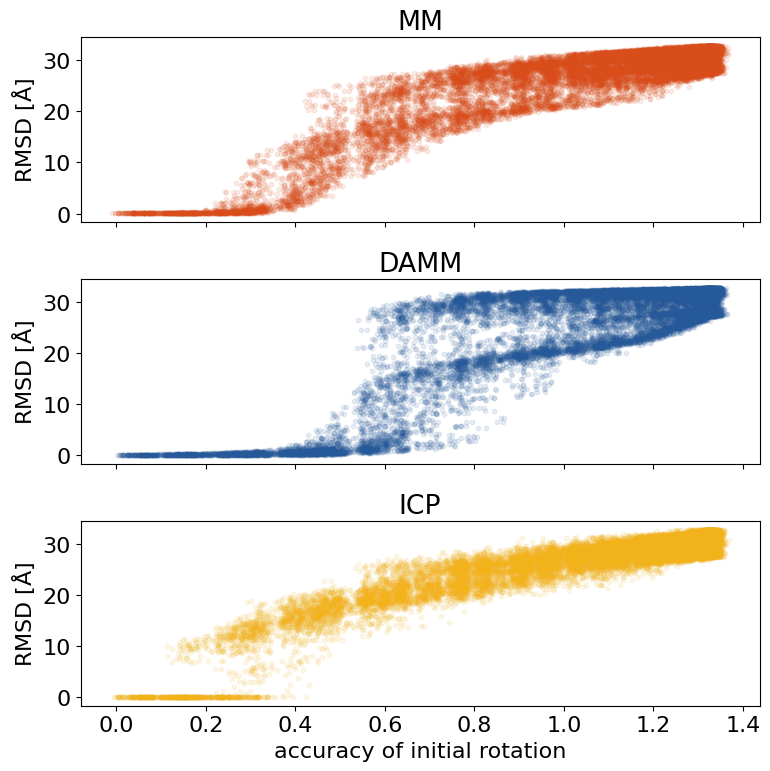}
  \caption{Radius of convergence of the local registration methods. The TIM barrel structure 6HF2 was matched against itself starting from initial rotations that form a tessellation of rotation space at a very fine level of discretization. The distance between the initial and the correct rotation is plotted against the RMSD achieved by each registration method. }\label{fig:radius-of-convergence}
\end{figure}
Figure \ref{fig:local} shows the performance of the local registration methods on the self-matching benchmark. We use the $\alpha$-recall of the RMSD defined in Eq. (\ref{eq:rmsd}) to assess the performance \citep{Zhou16}. The $\alpha$-recall is the fraction of tests on which a registration method reaches a final pose with RMSD below a given threshold $\alpha$. A good registration method should produce large fractions close to 100\%{} for small $\alpha$. On all test cases, the deterministic annealing approach performs best, reaching $\alpha$-recall near 100\%{} for small $\alpha$ below 1\AA. Also the MM approach without annealing outperforms ICP, but not as clearly as DAMM. All approaches face difficulties with target 6HF2, a member of the TIM barrel fold family. The likely reason for the problems with self-matching 6HF2 lies in the quasi-symmetry of the structure. Rotations about the barrel axis achieve similar kernel correlations and RMSDs, which adds to the severeness of the registration problem, because the chance of getting trapped in a local optimum is increased. Further details on the performance of the local optimization methods are given in Supplementary Tables S1 and S2. 

A systematic discretization of rotation space \citep{Straub17,vakili2021bayesian} allows us to estimate the radius of convergence of the three local registration algorithms. To do so, we match the TIM barrel structure 6HF2 against itself (no random permutation or transformation) such that the correct pose is $(\vec{I}, \vec{0})$. Using a tessellation of $SO(3)$ based on 21792 rotations $\R_n$, we run all three algorithms from initial poses $(\R_n, \vec{0})$. The performance of the registration algorithms is measured by the registration error \citep{Gao19}:
\begin{equation}\label{eq:error}
  \err(\R, \T) = \sqrt{\frac{1}{J} \sum_{j=1}^J \|\hat{\R}\Y_j + \hat{\T} - \R\Y_j - \T\|^2}
\end{equation}
where $(\hat{\R}, \hat{\T})$ is the ground truth pose (here $(\vec{I}, \vec{0})$). Notice that the registration error (\ref{eq:error}) differs from the RMSD defined in Eq. (\ref{eq:rmsd}), and can be interpreted as the mean squared deviation between the correctly transformed source and its pose found by a registration method. 

As is evident from Figure \ref{fig:radius-of-convergence}, the radius of convergence is highest for the MM methods, confirming our findings from the previous tests. The MM algorithms find the correct pose even, if the initial rotation has a distance smaller than 0.2 (MM) or 0.4 (DAMM) to the correct rotation. On the other hand, ICP starts to produce suboptimal fits already at rotations as close as 0.1. Similar results were obtained for the other target structures (see Supplementary Figures \ref{fig:convergence} and \ref{fig:convergence-rmsd}). 

\begin{figure}
  \centering
  \includegraphics[width=0.95\textwidth]{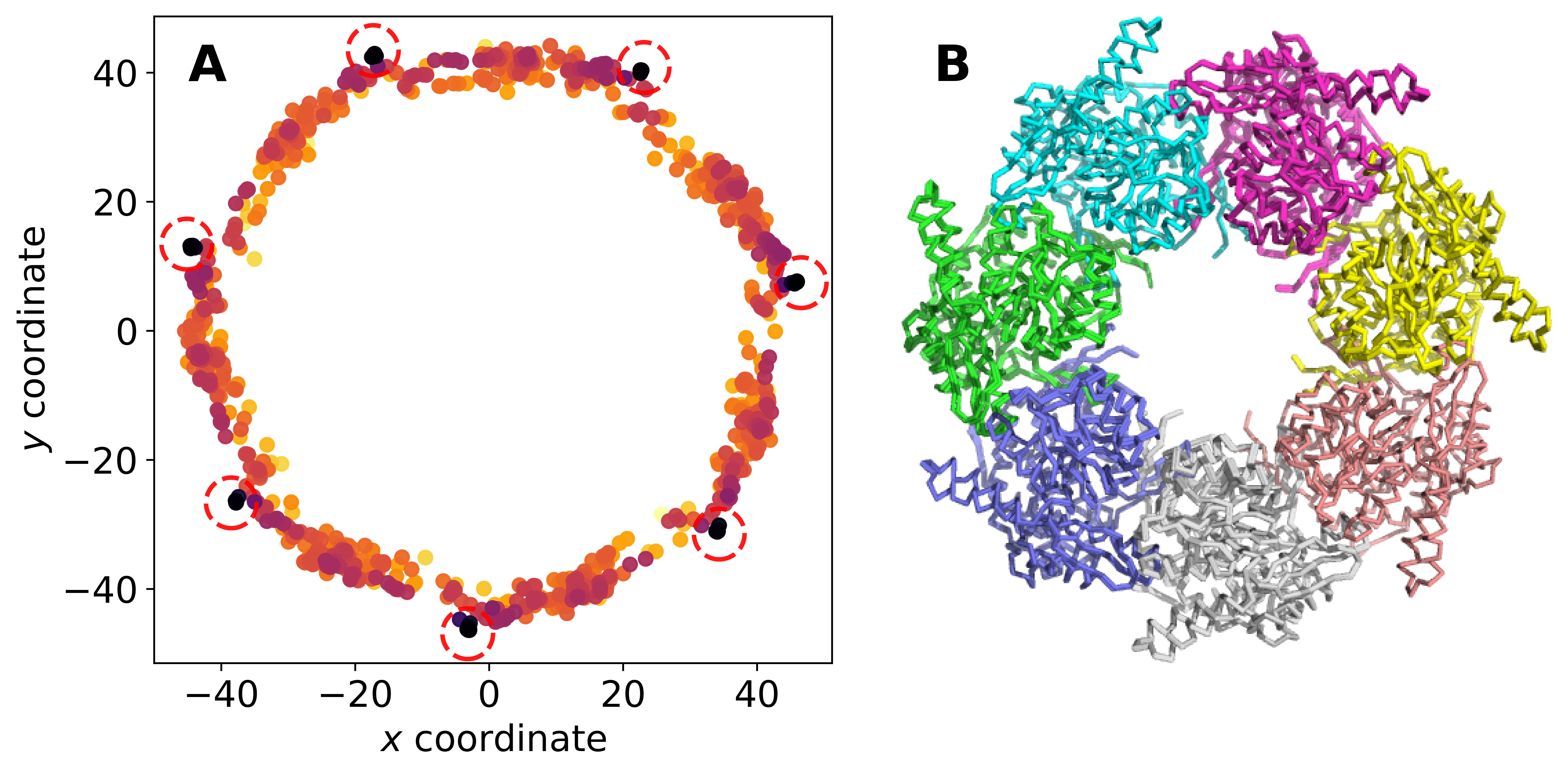}
  \caption{Global rigid registration by random search. \textbf{(A)} Center of mass of subunit A of GroEL projected along the symmetry axis for 1000 poses found by MM. The color encodes the kernel correlation of the pose (dark blue corresponds to a high kernel correlation, yellow corresponds to a bad fit). Dashed circles indicate the projected centers of mass of the subunits in the target (1OEL). \textbf{(B)} Ribbon representation of the top 50 poses found by MM. The color indicates poses that are superimposed onto the same subunit of the target (chain A to chain G from 1OEL). }\label{fig:groel}
\end{figure}
\subsection{Performance of global registration by random search}
Both the MM algorithms as well as ICP are local optimization methods and suffer from getting trapped in local minima. A simple strategy to locate the globally best fit is to run repeated optimizations starting from random initial poses. As a test case we consider the structure of GroEL which is composed of seven identical subunits exhibiting a 7-fold cyclic symmetry. The task is to superimpose a single subunit, subunit A, onto the GroEL ring structure. The kernel correlation of this superposition task has seven global minima corresponding to fitting one subunit onto any of the seven subunits in the target structure. In addition to the seven global optima, there are a multitude of local minima corresponding to partial matches between subunits. 

To tackle this challenging superposition task, we launch repeated MM runs from random rotations and translations. The initial translations are uniformly sampled from the bounding box of the target, the GroEL ring. The initial rotations are uniformly sampled over $SO(3)$. We first evaluate the grid-based kernel correlation for $10^5$ initial poses and keep the 1000 best initial poses achieving the highest kernel correlation. For each of the 1000 best initial poses, we then run the grid-based implementation of MM (see Supplementary Material \ref{sec:grid-MM}). These computations only take a few seconds on a standard notebook. 

Figure \ref{fig:groel}A shows center of mass of subunit A (the source) after applying the 1000 poses found by MM. Among the 1000 poses are very close (local) matches of subunit A onto each of the subunits in the target (see also Fig. \ref{fig:groel}B). But the large majority of poses correspond to a poor fit indicated by a low kernel correlation. This means that our global registration strategy indeed finds all global optima. However, to guarantee that we do not miss the globally best registration, we have to run MM multiple times from many initial poses. This is feasible thanks to the speed-up resulting from the grid approximation of the kernel correlation (see Section \ref{sec:grid}). 

\begin{figure}
  \centering
  \includegraphics[width=0.5\textwidth]{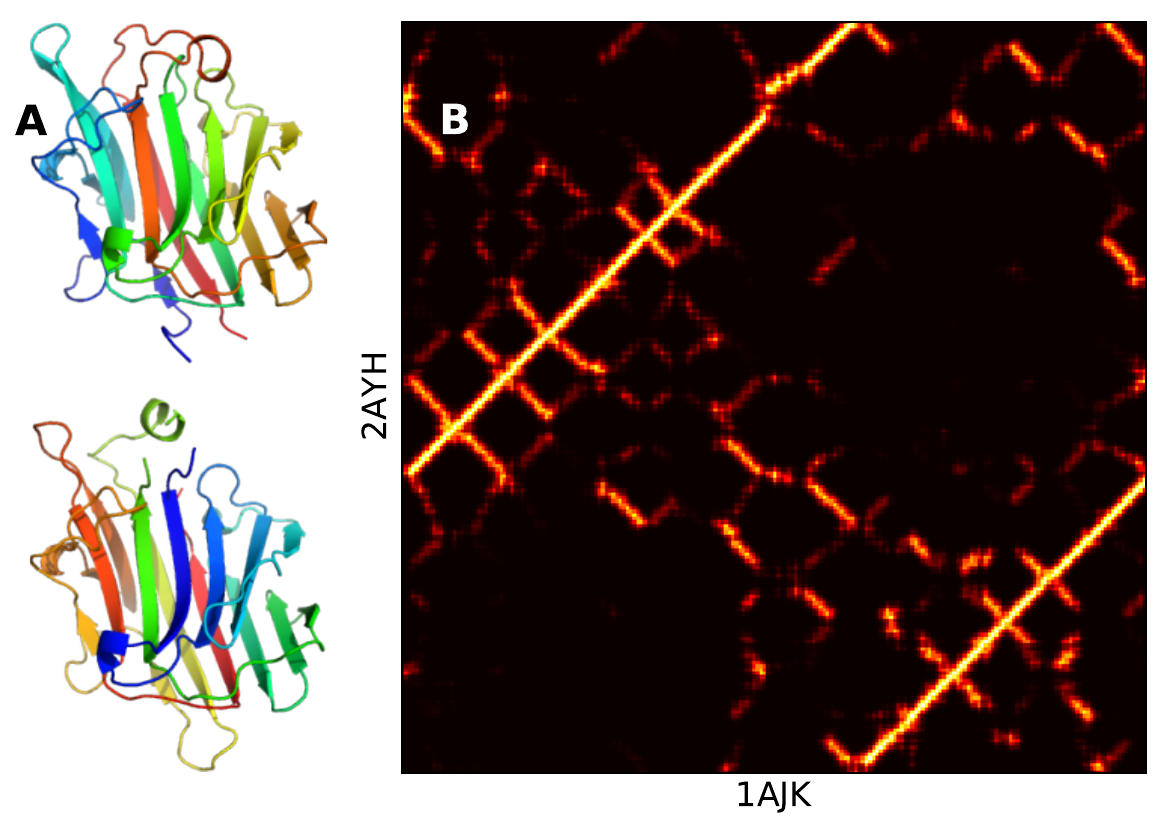}
  \caption{KC-based matching of circularly permuted structures 2AHY and 1AJK. \textbf{(A)} The left panel shows both structures after registration by maximizing the kernel correlation (Pymol's {\em chainbow} coloring indicates the order of amino acids in each structure). \textbf{(B)} The right panel shows the kernel matrix has a heatmap. The row indices correspond to the amino acid sequence of 2AYH running from bottom (N-terminus) to top (C-terminus). The column indices correspond to the sequence of 1AJK and run from left (N-terminus) to right (C-terminus). The brightness of the heatmap is directly proportional to the entries in the kernel matrix. }\label{fig:circular-permutation}
\end{figure}
\subsection{Comparison of circularly permuted structures}
Circular permutation breaks the sequential order of the amino acids and thereby poses a challenge to standard protein alignment methods. The kernel correlation does not require a position-to-position correspondence between both structures and is invariant under shuffling the order of points in each cloud. Therefore, circularly permuted structures can directly be superimposed and compared with KC-based registration methods.

To illustrate this point, we match the two circularly permuted structures 2AYH and 1AJK using the deterministic annealing approach with $\sigma=5$ \AA. The aligned structures are shown in Figure \ref{fig:circular-permutation}A. The correct alignment can be found very rapidly with DAMM. Out of 10 random initial poses, 3 produced the correct structural fit. At the right (Fig. \ref{fig:circular-permutation}B), we show the kernel matrix with elements $\phi_\sigma(\|\X_i - \R \Y_j - \T\|)$ as a heatmap. The kernel matrix clearly delineates corresponding structural regions, which are indicated by ``hot'' matrix elements that run parallel to the diagonal and are ``folded'' due to circular permutation. 

\subsection{Rigid fitting of bead models from small-angle scattering}
\begin{figure}%[hbt!]
  \centering
  \includegraphics[width=0.5\textwidth]{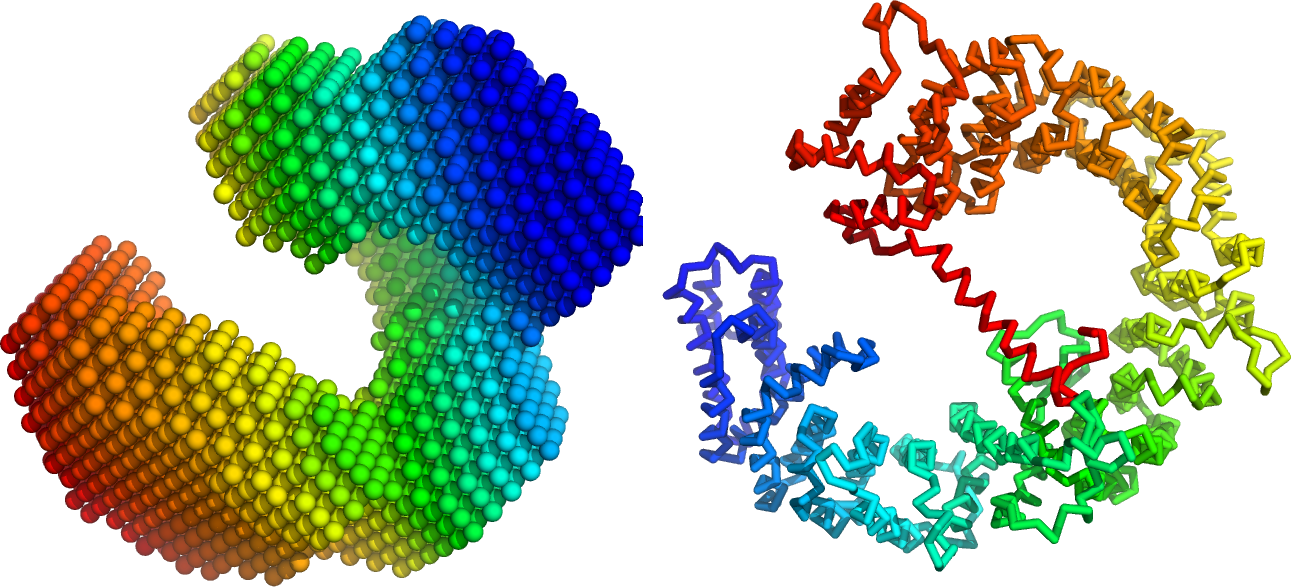}
  \caption{Docking a crystal structure of exporting CRM1 into a bead model derived from a SAS curve. The bead model obtained from SASBDB is shown on the left. The right panel shows the high-resolution structure 4HZK that was docked into the bead model by maximizing the kernel correlation. }\label{fig:SAS}
\end{figure}

Our registration algorithms can also be used to dock structures into bead models derived from small-angle scattering (SAS) curves. The bead models are obtained from the Small Angle Scattering Biological Data Bank (SASBDB) \citep{Valentini14}. The first target is a bead model of exportin CRM1 derived from a SAS curve (SASBDB code SASDAJ4). We fit the crystal structure 4HZK into the bead model by maximizing the kernel correlation with $\sigma = 5$ \AA. Again, we use the deterministic annealing approach to find the pose that maximizes KC. Figure \ref{fig:SAS} shows the SAS bead model and a CA trace of the superimposed crystal structure. The correlation between both point clouds is 75 \%. More examples can be found in Supplementary Figures S5 and S6. 

%

%%% Rigid docking into cryo-EM maps

\begin{figure}
  \centering
  \includegraphics[width=0.95\textwidth]{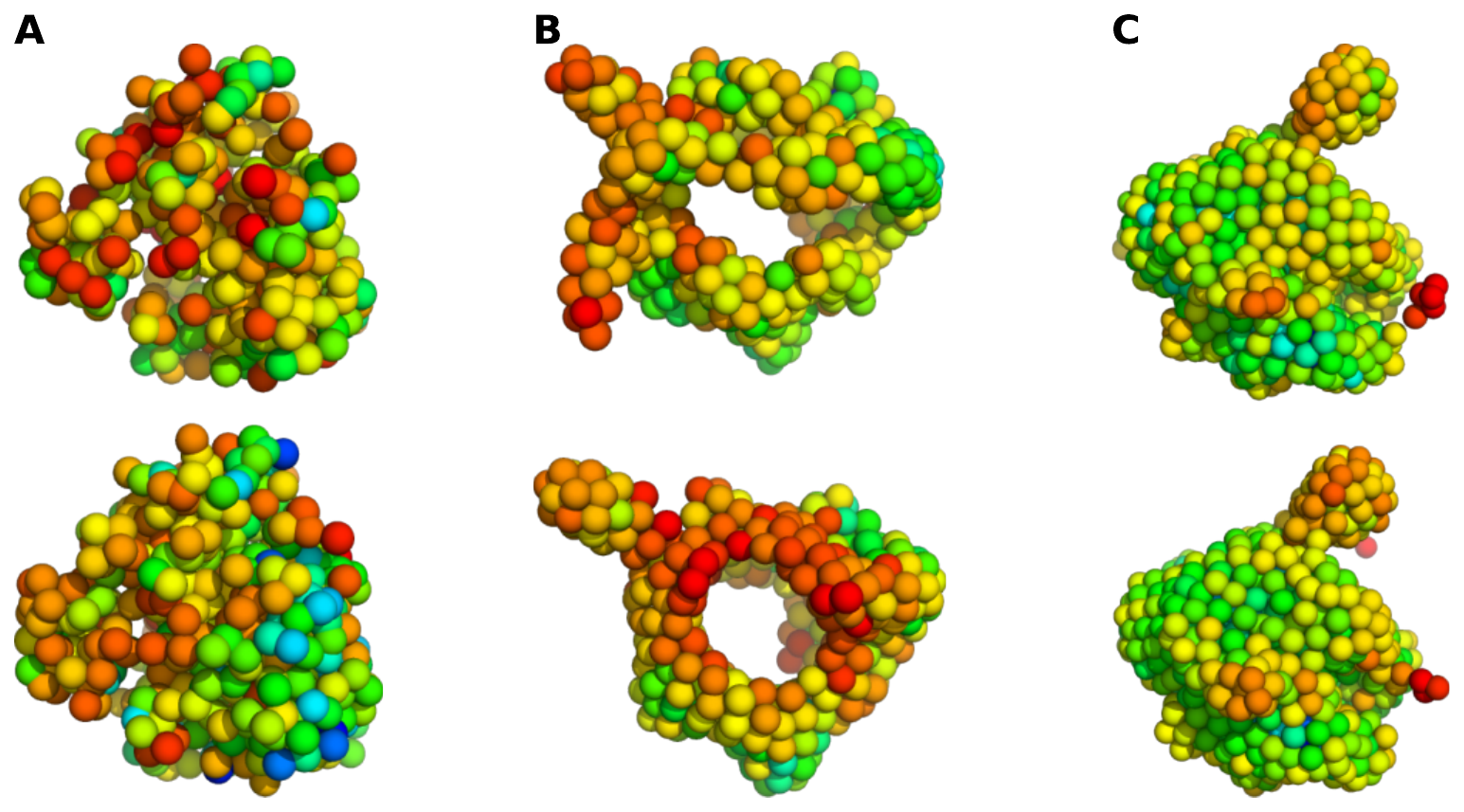}
  \caption{Superposition of low-resolution cryo-EM maps by rigid point cloud registration. The colors indicate the weight of the particles ranging from blue (high weight) to red (low weight). \textbf{(A)} Superposition of two bead models of intermediate-resolution maps of the 80S ribosome (EMD-1067, EMD-1343). \textbf{(B)} Two bead models of the free and nucleotide-bound structure of axonemal dynein-c (EMD-2155, EMD-2156). \textbf{(C)} Two bead models of human RNA polymerase II (EMD-2189, EMD-2190) in complex with different RNAs. } \label{fig:dock_emdb}
\end{figure}

\subsection{Rigid fitting of cryo-EM maps}
Next, we use the point-cloud registration methods to align two cryo-EM maps. We superimpose three pairs of low-resolution maps after converting them to weighted point clouds. All density maps are downloaded from the EMDataBank \citep{Lawson10} and converted to weighted point clouds by running the DP-means algorithm \citep{Kulis12}. We set the desired bead radius to 10 \AA{} for the first pair (two 80S ribosome maps) and to 5 \AA{} for the second and third pairs (characterizing a motor protein and RNA polymerase II). The first pair of medium-resolution maps shows the 80S ribosome at 11.7 \AA{} (EMD-1067) and 9.7 \AA{} resolution (EMD-1343). DP-means generates weighted point clouds with 425/666 beads representing EMD-1067/EMD-1343. The second docking task is to superimpose two low-resolution maps of axonemal dynein-c without and with nucleotide bound. The low-resolution maps EMD-2155 (apo dynein-c at 19 \AA{} resolution) and EMD-2156 (dynein-c with bound nucleotide at 22 \AA{} resolution) are represented by 433 and 405 beads, respectively. The third task is to superimpose bead models derived from EM maps EMD-2189 and EMD-2190 showing human RNA polymerase II at 25 \AA{} resolution in complex with different RNAs. The bead models are composed of 889 and 951 particles. In each of the docking tasks, the kernel bandwidth $\sigma$ was chosen to be twice as large as the bead radius.

Figure \ref{fig:dock_emdb} shows the superpositions obtained with the deterministic annealing approach. Visual inspection reveals that the 3D superpositions found by DAMM are meaningful. To quantify this further, we also investigated the correspondence between the negative log kernel correlation (which is the target function of the MM algorithms) and more traditional measures for assessing the overlap of two structures or 3D density maps. The kernel correlation can serve as a surrogate of the cross-correlation coefficient (CCC), which is typically maximized to superimpose two cryo-EM maps using a voxel representation. Supplementary Figure S7 shows that there is high correlation between both metrics. An advantage of the kernel correlation compared to the CCC is that it can be evaluated very efficiently and does not require the interpolation of the moving cryo-EM structure over a voxel grid. Similarly, we also see a high agreement between $-\log\KC$ and the RMSD as defined in Eq. (\ref{eq:rmsd}). 

\subsection{Rigid fitting of subunits into cryo-EM density maps of symmetric assemblies}
Finally, we use KC-based point cloud registration to dock a high-resolution structure of a single subunit into a cryo-EM map of a symmetric assembly. As is shown in the Supplementary Material, the kernel correlation as well as the majorization-minimization strategy for finding the best superposition can readily be adapted to the symmetric case.

We demonstrate the ability to dock structures into symmetric assemblies for the high-resolution map of the capsaicin receptor TRPV1 (EMD-5778). This assembly exhibits a four-fold cyclic symmetry and served as a target for rigidly fitting a single subunit into the map. The map and the symmetry operators were downloaded from the EMDataBank. We converted the map to a weighted point cloud by applying DP-means using a bead radius of 5 \AA. The modeled structure of a single subunit (PDB code 3J5P) was docked into the assembly by maximizing the symmetrized kernel correlation.  

Figure \ref{fig:symmetric} shows the cryo-EM map of the assembly and its point cloud representation next to the assembly predicted by fitting the structure of the subunit into the assembly. Visual inspection confirms that the subunit has been docked correctly into the point cloud representing the assembly. The correlation between both point clouds is 65 \%. More examples of rigid fits into symmetric assemblies are presented in Supplementary Figure S8. 

\begin{figure*}%[!tpb]
  \begin{center}
    \centering
    \includegraphics[width=\textwidth]{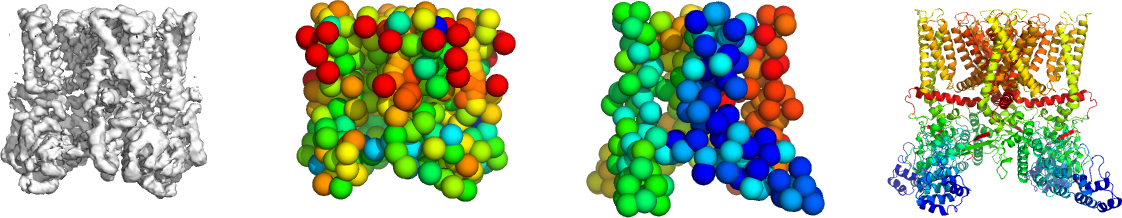}
    \caption{Rigid fitting of an atomic structure of a subunit into a high-resolution map of the symmetric channel TRPV1. Left: Cryo-EM map EMD-5778. Middle left: Bead model representing the cryo-EM map where the colors indicate the weight of particles. Middle right: Docked structure of the subunit and symmetry mates shown as bead models. Right: Docked high-resolution structure. } \label{fig:symmetric}
  \end{center}
\end{figure*}

\section{Conclusion}
3D superposition of biomolecular structure is a common task in structural biology that is typically solved by specialized algorithms and software that depend on the representation of a 3D structure. In this article, we address 3D fitting problems within a common framework based on the registration of weighted point clouds. As a measure of similarity, we use the kernel correlation, and introduce iterative algorithms for optimizing it so as to superimpose two point clouds. 

An advantage of the kernel correlation over RMSD-based approaches is that KC does not require a point-to-point correspondence. This advantage comes at the cost of having to deal with an objective function that exhibits multiple optima and is therefore more difficult to optimize than the standard RMSD or its modified versions. Local point-cloud registration algorithms therefore risk to get trapped in local optima. To overcome these challenges, we introduced an iterative MM algorithm and its annealed version, which both have a larger radius of convergence and thereby a lower chance of getting trapped in suboptima than the commonly used ICP method. Due to the generality of the representation, our rigid registration approach can be applied to various 3D fitting problems including the comparison of circularly permuted structures or the superposition of bead models and density maps from cryo-EM.  

There is still room to improve the efficiency of the registration algorithm. Its current implementation is not fast enough for large-scale similarity searches based on point clouds. The computation time to evaluate the kernel correlation scales with the size of both point clouds. Therefore, to improve the search over all rigid transformation, we plan to pursue a multiscale approach based on a hierarchical representation of point clouds. Coarser representations would involve a smaller number of points thereby allowing us to evaluate the kernel correlation more rapidly. Combined with a global grid search, a multiscale approach should enable us to exhaustively scan all rigid transformations and further reduce the chance to miss the global optimum. Another interesting direction is to fit multiple subunits into a point cloud representing an assembly, and to fit 3D point clouds against 2D point clouds derived from projection images obtained with electron microscopy or tomography as well as other imaging methods. 

\section*{Funding}
MH, AK and NV gratefully acknowledges funding by the Carl Zeiss Foundation within the program ``CZS Stiftungsprofessuren''.  MH and NV acknowledge funding from the German Research Foundation (DFG) within SFB 860 subproject B09. MH and AK are grateful for the support of the DFG within project 432680300 - SFB 1456 subprojects A05.

%\bibliographystyle{natbib}
%\bibliography{references.bib}

\clearpage

\begin{appendix}
  
\renewcommand\thefigure{S\arabic{figure}}    
\setcounter{figure}{0}    

\renewcommand\thetable{S\arabic{table}}    
\setcounter{table}{0}    

\section{Algorithms}
This section lists all algorithms presented in the paper.
\subsection{MM algorithm to locally minimize the negative log kernel correlation}
To optimize the kernel correlation locally, we run the iterations (with iteration index $n$ used as upperscript) starting from an initial pose $(\R^{(0)}, \T^{(0)})$: 
\begin{align}\label{eq:mm}
  & w_{ij}^{(n)} \leftarrow \frac{\phi_\sigma(\|\X_i - \R^{(n)} \Y_j - \T^{(n)}\|)}{ \sum_{i'j'}  q_{i'} p_{j'} \phi_\sigma(\|\X_{i'} - \R^{(n)} \Y_{j'} - \T^{(n)}\|)}  \\[0.5cm]
  & (\R^{(n+1)}, \T^{(n+1)}) \leftarrow \underset{\R\in SO(3),\, \T\in\mathbb R^3}{\text{argmin}} \upperbound^{(n)}(\R, \T)
\end{align}
where
\begin{equation}\label{eq:upper_bound}
  \upperbound^{(n)}(\R, \T) = \frac{1}{2} \sum_{ij} q_i p_j w^{(n)}_{ij} \|\X_i - \R \Y_j - \T\|^2
\end{equation}
is an upper bound (up to a constant) of $-\log\KC$. 

\subsection{Deterministic annealing during MM}
In deterministic annealing, we decrease the bandwidth $\sigma$ during the MM iterations according to a schedule $\sigma^{(0)} > \sigma^{(1)} \ge \sigma^{(2)} > \ldots$: 

\begin{align}\label{eq:mm-anneal}
  & w_{ij}^{(n)} \leftarrow \frac{\phi_{\sigma^{(n)}}(\|\X_i - \R^{(n)} \Y_j - t^{(n)}\|)}{ \sum_{i'j'}  q_{i'} p_{j'} \phi_{\sigma^{(n)}}(\|\X_{i'} - \R^{(n)} \Y_{j'} - \T^{(n)}\|)}  \\[0.5cm]
& (\R^{(n+1)}, \T^{(n+1)}) \leftarrow \underset{\R\in SO(3),\, \T\in\mathbb R^3}{\text{argmin}} \upperbound^{(n)}(\R, \T)
\end{align}

\subsection{Iterative closest point (ICP)}
Iterative closest point (ICP) is among the most commonly used approaches for rigid registration of point clouds. ICP iterates over updating a point-to-point correspondence $i(j)$ between points in the target $(\vec{X}\!, \vec{q})$ and the source $(\vec{Y}\!, \vec{p})$, and updating the rigid transformation $(\R, \T)$: 

\begin{align}\label{eq:icp}
& i^{(n)}(j) = \underset{i\in \{1, \ldots, I\}}{\text{argmin}}\,\, \|\X_i - \R^{(n)} \Y_j - \T^{(n)}\| \\[0.5cm]
& (\R^{(n+1)}, \T^{(n+1)}) \leftarrow \underset{\R\in SO(3),\, \T\in\mathbb R^3}{\text{argmin}} \,\, \sum_{j} \|\X_{i^{(n)}(j)} - \R \Y_j - \T\|^2
\end{align}

\subsection{Minimization of the upper bound}
To minimize $\upperbound^{(n)}$, we use a variant of the Kabsch algorithm. The gradient of the upper bound with respect to the translation is
\begin{equation}\label{eq:upper_bound_gradient_t}
  \vec\nabla_{\T} \upperbound^{(n)}(\R, \T) 
  = \T - \sum_{ij} q_i p_j w^{(n)}_{ij} (\X_i - \R \Y_j) 
  = \T - \underbrace{\sum_{ij} q_i p_j w^{(n)}_{ij} \X_i}_{\overline{\X}} + \R \underbrace{\sum_{ij} q_i p_j w^{(n)}_{ij} \Y_j}_{\overline{\Y}} 
\end{equation}
Setting $\nabla_{\T} \upperbound^{(n)}$ to zero and solving for $\T$ yields the optimal translation as a function of the rotation: 
\begin{equation}\label{eq:t_estimate}
  \hat{\T} = \overline{\X} - \R\overline{\Y}
\end{equation}
Plugging this estimator into $\upperbound^{(n)}$ yields an upper bound that depends only on $\R$: 
\begin{align}\label{eq:upper_bound_rotation}
  \upperbound^{(n)}(\R, \vec{\hat t}) 
  &= \frac{1}{2} \sum_{ij} q_i p_j w^{(n)}_{ij} \|(\X_i - \overline{\X}) - \R (\Y_j - \overline{\Y})\|^2 \\
  &= \frac{1}{2} \sum_{ij} q_i p_j w^{(n)}_{ij} \left(\|\X_i - \overline{\X}\|^2 +  \|\Y_j - \overline{\Y}\|^2 \right) - 
  \text{tr}\biggl(\underbrace{\sum_{ij} q_i p_j w^{(n)}_{ij} (\Y_j - \overline{\Y})(\X_i - \overline{\X})^T}_{\vec{S}^T} \R\biggr)
\end{align}
To minimize $\upperbound^{(n)}$ we have to {\em maximize} $\text{tr}(\pmb S^T \R)$ which can be achieved by computing a singular value decomposition of the $3\times 3$ matrix $\pmb S$ as in the standard Kabsch algorithm. 

\subsection{Data structures for fast evaluation of the kernel correlation}
We use several approximations to speed up the evaluation of $-\log\KC$. 
\subsubsection{Distance cutoff}
The support of the RBF kernel $\phi_\sigma$ is the entire positive real axis, but for $r > \nsigma{} \sigma$ the contributions are negligible. Therefore, we set
\begin{equation}\label{eq:cutoff}
  \phi_\sigma(r) = \begin{cases}
    \frac{1}{(2\pi\sigma^2)^{3/2}} \exp\left\{-\frac{r^2}{2\sigma^2} \right\} & 0 \le r < \nsigma{}\sigma \\
0 & r \ge \nsigma{} \sigma
  \end{cases}
\end{equation}
This approximation reduces the evaluation of $\|\pmb x_i - \pmb R \pmb y_j - \pmb t\|$ to nearest neighbors inside a ball of radius $\nsigma{}\sigma$. There exist efficient data structures for nearest-neighbor searches such as k-d trees, ball trees and neighbor lists.

\subsubsection{Gridding}
The kernel density estimates (KDEs) induced by both point clouds are
\begin{equation}\label{eq:kde}
  q_\sigma(\pmb x) = \sum_{i=1}^I q_i\, \phi_\sigma(\|\pmb x - \pmb x_i\|), \quad{}
  p_\sigma(\pmb x) = \sum_{j=1}^J p_j\, \phi_\sigma(\|\pmb x - \pmb y_j\|)
\end{equation}
We can interpret the kernel correlation as the scalar product of two KDEs:
\begin{equation}\label{eq:corr_inner}
\KC(\pmb R, \pmb t) = \langle q_{\sigma_1}, p_{\sigma_2}(\pmb R^T(\cdot - \pmb t)) \rangle
\end{equation}
where the bandwidths of the KDEs need to satisfy $\sigma^2 = \sigma_1^2 + \sigma_2^2$. We can use a regular cubic grid
\begin{equation}\label{eq:cubic_grid}
\mathcal G = \left\{\pmb x_0 + \Delta (n_1, n_2, n_3)^T \mid n_1 \in [N_1], n_2 \in [N_2], n_3 \in [N_3] \right\}
\end{equation}
to approximate the integral in the scalar product. Here, $[N] := \{0, 1, \ldots, N-1\}$ for some natural number $N\in \mathbb N$. The scalar $\Delta > 0$ is the grid spacing and $\X_0 \in \mathbb R^3$ the grid origin. We then have
\begin{equation}\label{eq:gridding}
  \KC(\R, \T) = \int q_{\sigma_1}(\Z)\, p_{\sigma_2}(\pmb R^T(\Z - \T))\, \dd\Z
  \approx \sum_{\Z \in \mathcal G} q_{\sigma_1}(\Z)\, p_{\sigma_2}(\Z)\, .
\end{equation}
We now choose $\sigma_1=\sigma$ (bandwidth of the fixed target) and $\sigma_2=0$ (bandwidth of the moving source), then the source is represented by a sum of delta peaks
$$
p_0(\pmb x) = \sum_{j=1}^J p_j\, \delta(\|\pmb x - \pmb y_j\|)\, .
$$
In practice, the source density $p_0$ is obtained by the following algorithm:
\begin{itemize}
\item For a given pose $(\R, \T)$, transform the source $\Y_j \mapsto \R \Y_j + \T$.
\item Compute the multi-index of the grid cell that contains the transformed point $\Y_j$:
  \[
  \vec{n}_j = \text{round}((\R\Y_j + \T - \X_0) / \Delta) \in \mathbb N^3
  \]
\item Map the multi-index $\pmb n_j$ to a flat grid index
  $$
  n_j = N_2 N_3 n_{j1} + N_3 n_{j2} + n_{j3}
  $$
  enumerating all grid cells and add the weight $p_j$ of source point $\Y_j$ to grid cell $n_j$. 
\end{itemize}
The approximate kernel correlation can then be computed by
\begin{equation}\label{eq:gridding2}
  \kappa(\pmb R, \pmb t) \approx \sum_{\pmb x\in \mathcal G} q_\sigma(\pmb x)\, p_0(\pmb R^T(\pmb x - \pmb t)) = \sum_k q_k p_k 
\end{equation}
where $\{q_\sigma(\pmb x)\}_{\X\in\mathcal G} = (q_1, \ldots, q_{|\mathcal G|})$ is a size $|\mathcal G|$ vector that only needs to be computed once and $p_k$ is a sparse binary vector that can be computed very rapidly for a new pose. To compute $p_k$, the transformed source $\pmb R \pmb y_j + \pmb t$ is mapped to the grid $\mathcal G$ by applying the following steps:
\begin{enumerate}
\item $\pmb z_j \gets \pmb R \pmb y_j + \pmb t$ (transform source)
\item $\pmb z_j \gets \pmb z_j - \pmb x_0$ (subtract grid origin)
\item $\pmb z_j \gets \text{round}(\pmb z_j / \Delta)$ (map to multi-index)
\end{enumerate}
This runs in $\mathcal O(J)$. 

\clearpage

\section{Computation times for fast evaluation of the kernel correlation}
The kernel correlation is evaluated on two large point clouds derived from PDB files 5M52 and 5M5P where now each listed atom (not only carbon alphas) defines a point. The target point cloud comprises 34512 and the source 67309 points. The evaluation of the exact kernel correlation without a hard distance cutoff on $\phi$ requires prohibitively long computation times and/or a large working memory (our default implementation of KC computes the full distance matrix between all pairs of points in the target and source). As a reference to compare different implementation of the kernel correlation, we therefore use a version based on a kernel density estimate (KDE) from scikit-learn. The KDE-based implementation took $650 \pm 19$ sec. per pose for $\sigma=3$ \AA{} on a i7 processor (1.8 GHz oct-core using a single thread only) on 10 random poses. By using the implementation described in section 2.3, the time for computing the kernel correlation is sped up significantly by 1 to 3 orders of magnitude as shown in the figure below. The fast KC implementations are based on approximations ($3\sigma$ cutoff and/or discretization), which result in approximate KC values. However, the accuracy of the approximate KC values is quite high as indicates by the Pearson correlation coefficients approximate and exact KC values: 100.00 \%{} ($k$-d tree), 100.00 \%{} (ball tree), and 99.98\%{} (cubic grid). 
\\

\begin{figure}[hbt!]
  \centering
  \includegraphics[width=0.75\textwidth]{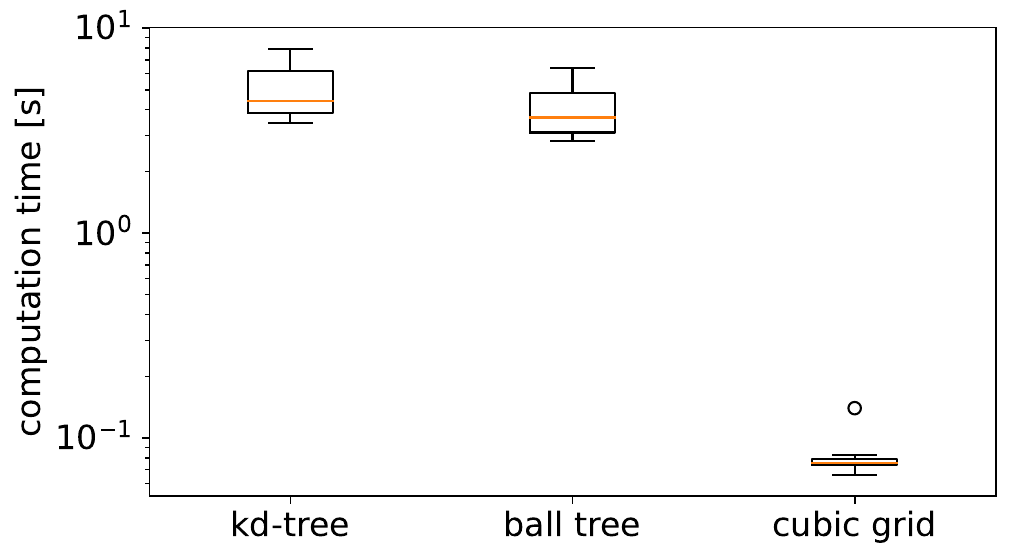}
  \caption{Computation times for evaluating the kernel correlation with an implementation using $k$-d trees, ball trees, and a cubic grid.
  }\label{fig:times}
\end{figure}

\clearpage

\section{Iterative majorization-minimization (MM)}
The figure below illustrates the majorization-minimization (MM) approach used in this paper to minimize the negative log kernel correlation $-\log\KC$. Instead of minimizing $-\log\KC$ directly over all rotational and translational degrees of freedom, our MM algorithms minimize an upper bound $\upperbound^{(n)}$ detailed in Eq. (\ref{eq:upper_bound}) (also see Eq. (7) in the main article). The progression of the MM iterations is shown for a self-matching task where PDB structure 6JC2 is fitted against itself. As is evidenced by the plot, the upper bound $\upperbound^{(n)}$ is quite tight and becomes even tighter in the course of the MM iterations as $n$ increases. \\

\begin{figure}[hbt!]
  \centering
  \includegraphics[width=0.75\textwidth]{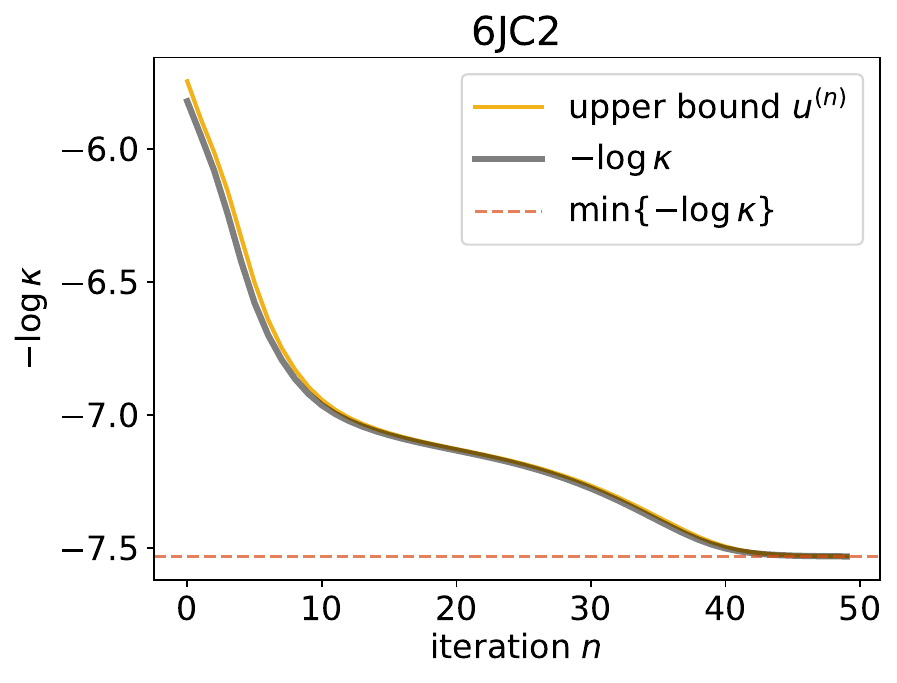}
  \caption{Evolution of the negative log kernel correlation $-\log\KC$, our objective function to superimpose two weighted point clouds, during the MM iterations. Instead of $-\log\KC$ itself, each iteration minimizes the upper bound $\upperbound^{(n)}$ shown in yellow. }\label{fig:upper_bound}
\end{figure}

\clearpage

\section{Majorization-minimization (MM) of grid-based kernel correlation}\label{sec:grid-MM}
The MM update also benefit from a speed up based on approximating the kernel correlation on a regular cubic grid. The weights of the current pose $(\R, \T)$ are %
\[
w_{ij} = \frac{\phi_\sigma(\|\X_i - \R \Y_j - \T\|)}{\sum_{i'j'} \phi_\sigma(\|\X_{i'} - \R \Y_{j'} - \T\|)}\, .
\]
The grid approximation of the Gaussian kernel is
\[
\phi_\sigma(\|\X_i - \R \Y_j - \T\|) \approx \sum_{\Z_k \in \mathcal G} \phi_\sigma(\|\X_i -  \Z_k\|) \delta(\|\Z_k - \R\Y_j - \T\|) = \sum_k \phi_{ik} \delta_{jk}
\]
where $\delta$ is the delta function and $\mathcal G$ a regular cubic grid whose grid cells are centered at $\Z_k$ and indexed by the flat index $k$. The kernel matrix $\phi_{ik}$ does not change in the course of the MM procedure.

In the MM iterations, we need to compute the weighted means $\overline{\X}$ and $\overline{\Y}$ of the target and the source. We have
\begin{equation}
\overline{\X}
= \sum_{ij} w_{ij} q_i p_j \X_i 
\approx \frac{\sum_{ijk} \phi_{ik} \delta_{jk} q_i p_j \X_i}{\sum_{ijk} \phi_{ik} \delta_{jk} q_i p_j} 
= \frac{\sum_k \left(\sum_i \phi_{ik} q_i \X_i\right) \left(\sum_j p_j \delta_{jk}\right)}{\sum_k \left(\sum_i q_i \phi_{ik}\right) \left(\sum_j p_j \delta_{jk} \right)} = \frac{\sum_k \tilde{\X}_k \tilde{p}_k}{\sum_k \tilde{q}_k \tilde{p}_k}
\end{equation}
where we defined $\tilde{q}_k = \sum_i q_i \phi_{ik}$ (the target KDE evaluated on the grid $\mathcal G$), $\tilde{p}_k = \sum_j p_j \delta_{jk}$ (the weighted sum of delta peaks located at the transformed source positions $\R\Y_j + \T$ evaluated on $\mathcal G$ via rounding), and $\tilde{\X}_k = \sum_i q_i \phi_{ik} \X_i$. All $\tilde{q}_k$ and $\tilde{\X}_k$ can be computed before launching the MM iterations and will stay constant in the course of the iterations. The gridded sum of delta functions $\tilde{p}_k$ has to be recomputed each time the source adopts a new pose, but computation of $\tilde{p}_k$ is very fast because it can be done by rounding the coordinates of the source points to grid cells. Moreover, $\tilde{p}_k$ is a sparse array and sums involving $\tilde{p}_k$ can be restricted to those grid cells that are occupied by a source point.

Similarly, we can show that
\begin{equation}
\overline{\Y}
= \sum_{ij} w_{ij} q_i p_j \Y_j 
\approx \frac{\sum_k \tilde{\Y}_k \tilde{q}_k}{\sum_k \tilde{q}_k \tilde{p}_k}
\end{equation}
with $\tilde{\Y}_k = \sum_j p_j \delta_{jk} \Y_j$ which is a sparse array of 3D vectors. Estimation of the optimal rotation is then achieved by computing the SVD of the matrix
\begin{align}
  \pmb S &= \sum_{ij} q_i p_j \phi_\sigma(\|\X_i - \R \Y_j - \T\|) (\X_i - \overline{\X}) (\Y_j - \overline{\Y})^T \\
  &\approx \sum_k \biggl(\sum_i q_i \phi_{ik} (\X_i - \overline{\X}) \biggr) \biggl(\sum_j p_j \delta_{jk} (\Y_j - \overline{\Y}) \biggr)^T \\
  &= \sum_k (\tilde{\X}_k - \tilde{q}_k \overline{\X}) (\tilde{\Y}_k - \tilde{p}_k \overline{\Y})^T\, .
\end{align}
Again, since $\tilde{p}_k$ and $\tilde{\Y}_k$ are sparse, the computation of $\pmb S$ is very fast. 

\clearpage

\section{Self-match benchmark with random initial rotation and translation}
Both in terms of the correlation (which is proportional to the objective function of MM and DAMM) as well as the RMSD (which is the objective function of ICP), the deterministic annealing approach performs best, reaching correlations near 100\%{} and RMSDs close to zero for most test cases: \\

\begin{table}[h!]
  \begin{center}
  \begin{tabular}{r c c c c c}
    \toprule
    & 6JC2 & 6HF2 & 1OEL & 5G5D & 6R4S \\
    \midrule
    MM & $\pmb{1.00 \pm 0.01}$ & $0.96 \pm 0.04$ & $0.96 \pm 0.08$ & $0.99 \pm 0.01$ & $0.99 \pm 0.03$ \\
    DAMM & $\pmb{1.00 \pm 0.00}$ & $\pmb{0.99 \pm 0.03}$ & $\pmb{1.00 \pm 0.02}$ & $\pmb{1.00 \pm 0.00}$ & $\pmb{1.00 \pm 0.01}$ \\
    ICP & $0.99 \pm 0.03$ & $0.92 \pm 0.05$ & $0.92 \pm 0.11$ & $0.98 \pm 0.02$ & $0.98 \pm 0.05$ \\
    \bottomrule
  \end{tabular}\caption{Average correlation coefficient.}
  \end{center}
\end{table}

\begin{table}[h!]
  \begin{center}
  \begin{tabular}{r c c c c c}
    \toprule
    & 6JC2 & 6HF2 & 1OEL & 5G5D & 6R4S \\
    \midrule
MM & $0.21 \pm 0.57$ & $2.14 \pm 1.49$ & $1.36 \pm 1.80$ & $0.64 \pm 1.03$ & $0.52 \pm 1.12$ \\
DAMM & $\pmb{0.00 \pm 0.02}$ & $\pmb{0.96 \pm 1.22}$ & $\pmb{0.07 \pm 0.49}$ & $\pmb{0.19 \pm 0.34}$ & $\pmb{0.04 \pm 0.36}$ \\
ICP & $0.28 \pm 0.96$ & $2.76 \pm 1.80$ & $1.96 \pm 2.35$ & $1.02 \pm 1.40$ & $0.48 \pm 1.34$ \\
    \bottomrule
  \end{tabular}\caption{Average RMSD in \AA.}
  \end{center}
\end{table}

Also the MM approach without annealing outperforms ICP, but not as clearly as DAMM. Nevertheless, all approaches face difficulties with target 6HF2, a member of the TIM barrel fold family. These difficulties can be overcome by using a larger number of initial poses. When we increased the number of random initial poses from 10 to 50, the average correlations improved to 99.9\%{} (MM), 100.0\% (DAMM), and 98.5\% (ICP), as do the average RMSDs: 0.2 \AA{} (MM), 0.1 \AA{} (DAMM), and 0.6 \AA{} (ICP). 

\clearpage

\section{Radius of convergence of iterative registration methods}
\begin{figure}[hbt!]
  \centering
  \includegraphics[width=\textwidth]{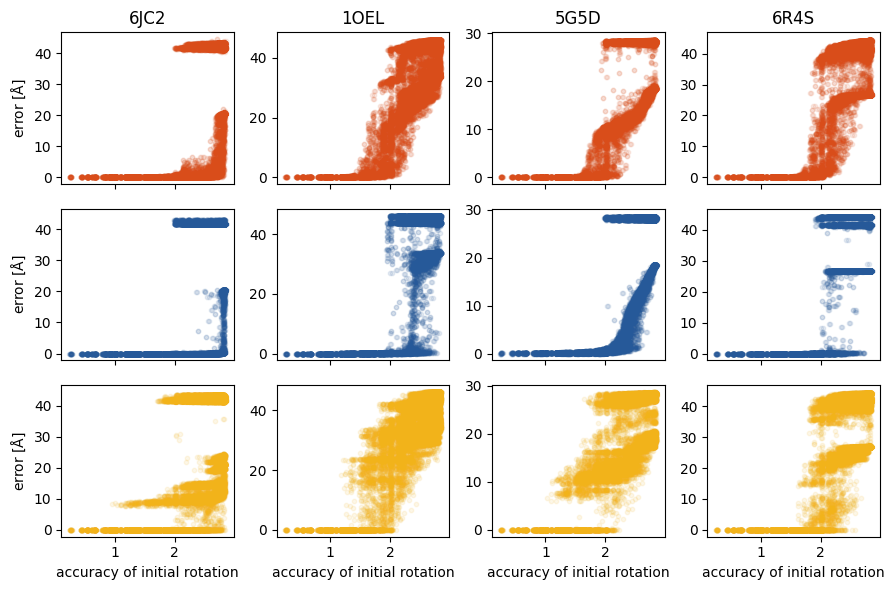}
  \caption{Radius of convergence. Distance between initial and true rotation versus the RMSD of the final pose generated by the registration methods: MM (red), DAMM (blue), ICP (yellow).}\label{fig:convergence}
\end{figure}

\begin{figure}[hbt!]
  \centering
  \includegraphics[width=\textwidth]{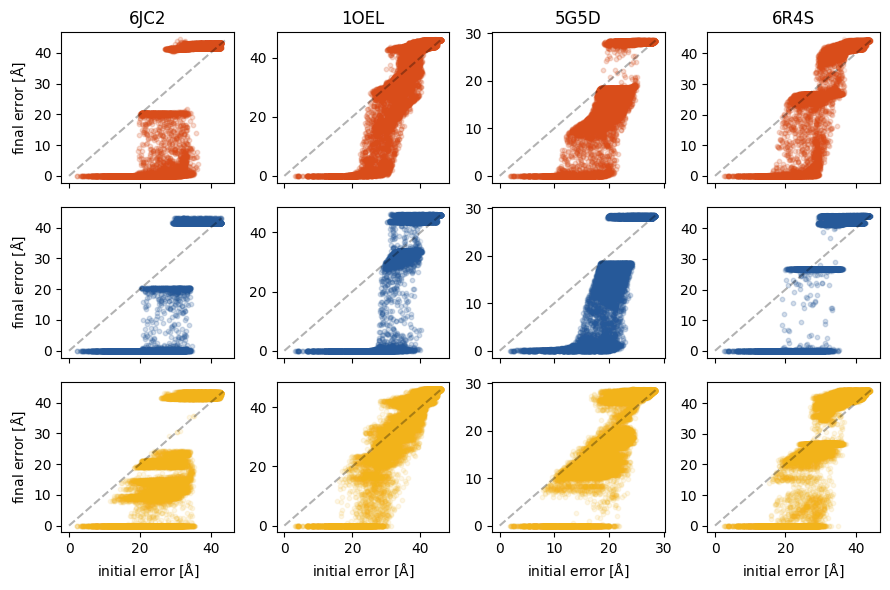}
  \caption{Radius of convergence. RMSD of initial pose versus RMSD of the final pose generated by the registration methods: MM (red), DAMM (blue), ICP (yellow).}\label{fig:convergence-rmsd}
\end{figure}

In addition to the example shown in Figure 2 of the main manuscript, we also assessed the radius of convergence of three registration methods (ICP: iterative closest point; MM: iterative majorization-minimization; DAMM: MM with deterministic annealing) using four other test cases. The setup is the same as described in the main paper. In all tests, the MM methods have a larger radius of convergence than ICP (see Fig. \ref{fig:convergence} and Fig. \ref{fig:convergence-rmsd}). \\

\clearpage

\clearpage

\begin{figure}[hbt!]
  \subfigure[\centering 6JC2]{
    \includegraphics[width=0.95\textwidth]{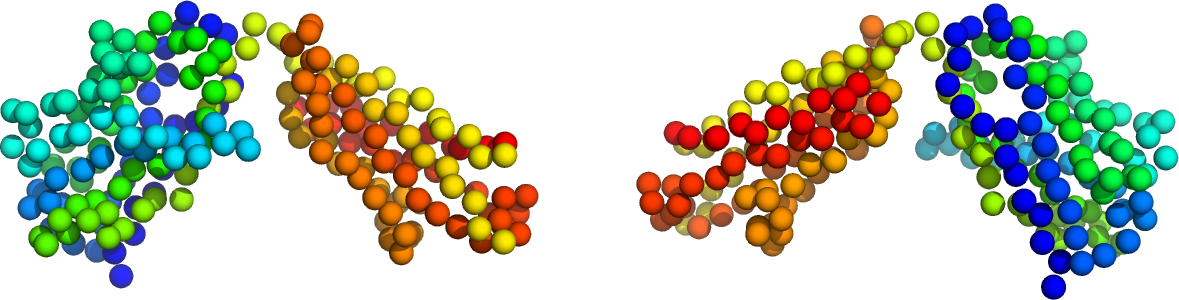}}
  \qquad
  \subfigure[\centering 5G5D]{
    \includegraphics[width=0.95\textwidth]{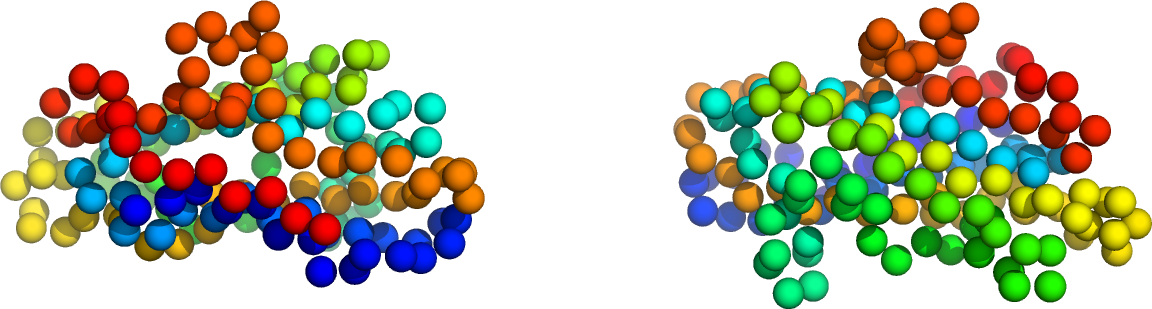}}
  \caption{Left: target point cloud. Right: bad pose with a good correlation but high RMSD.}\label{fig:docking}
\end{figure}

\noindent In most examples, we see the emergence of a second dominant cluster of solutions with high RMSD values. These registrations correspond to an alternative 3D superpositions due to a quasi-symmetry of the point cloud. For example, in case of 6JC2 we are dealing with a heterodimer where the two monomers are very similar to each other. The bad fit with an RMSD of 41.6 \AA{} achieves a correlation 91\%{} and is shown in Fig. \ref{fig:docking}(a). In case of 5G5D, we are dealing with a member of the Carbohydrate-binding domain superfamily, which also exhibits a quasi-symmetry. If we ignore the sequence information and only look at the spatial arrangement of CA atoms, the bad fit with an RMSD of 28.2 \AA{} achieves a correlation of 97\%{} and is shown in Fig. \ref{fig:docking}(b). \\

\clearpage

\section{Self-match benchmark with random initial rotation and grid search of the translation}
The self-matching benchmark (subsection 3.1.1 of the main article) was modified as follows. Instead of choosing the initial translation randomly, it was optimized by maximizing the kernel correlation over a regular cubic grid. The resulting average correlation values and RMSDs can found in the following figure:

\begin{figure}[hbt!]
  \centering
  \includegraphics[width=\columnwidth]{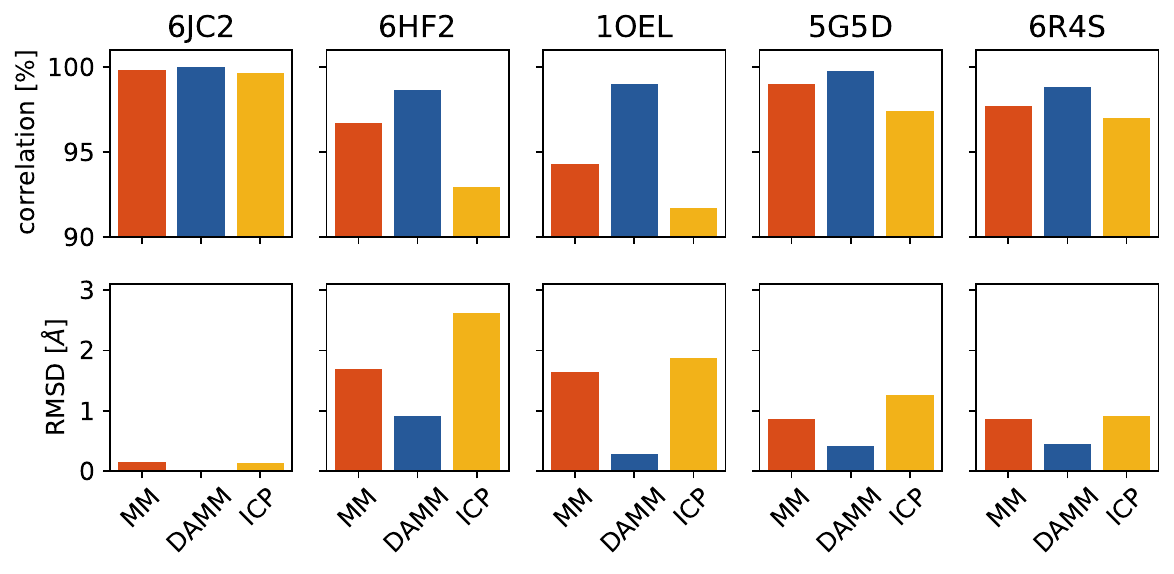}
  \caption{Testing the impact of using an initial translation that is optimized over a cubic grid of cell size 1 \AA. A PDB structure (indicated in panel titles) is fitted against a permuted and randomly transformed version of itself. The top row shows the correlation coefficient (ratio of actual kernel correlation and maximum achievable kernel correlation) obtained when starting local optimization runs from 10 random initial rotations (and optimized translations). The bottom row shows the RMSD (defined in equation 7 of the main paper) of corresponding points after striking the estimated pose. }
\end{figure}

\clearpage

\section{Fitting of atomic structures into bead models derived from small-angle scattering curves}
\begin{figure}[hbt!]
  \centering
  \includegraphics[width=0.9\textwidth]{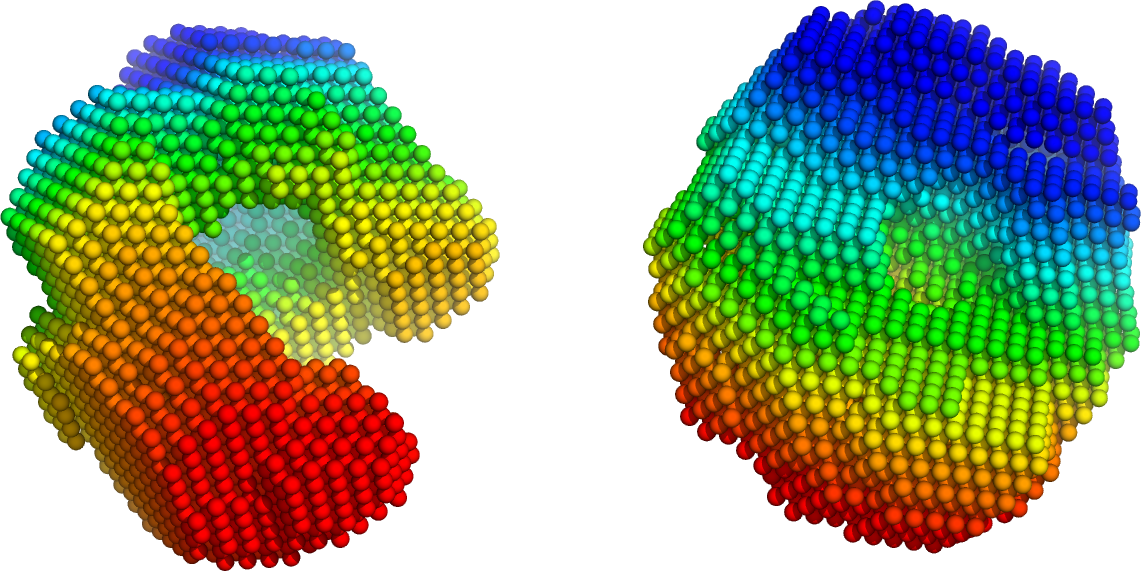}
  \caption{3D superposition of SASDAK4 (CRM1 RanGTP, shown on the right) onto SASDAJ4 (free CRM1, shown on the left). }
  \label{fig:SASDAJ4}
\end{figure}

The main article illustrates the ability to fit high-resolution structures into bead models from small-angle scattering (SAS) curves for an exportin structure. Here, we demonstrate this for two additional examples. The first example also involves exportin CRM1. We fitted a bead model of free CRM1 (SASDAJ4) against CRM1 RanGTP (SASDAK4). The figure \ref{fig:SASDAJ4} shows the final superposition found by maximizing the kernel correlation with $\sigma = 5$ \AA. The correlation of the final fit is 80.3\%. 

\begin{figure}[hbt!]
  \centering
  \includegraphics[width=0.5\textwidth]{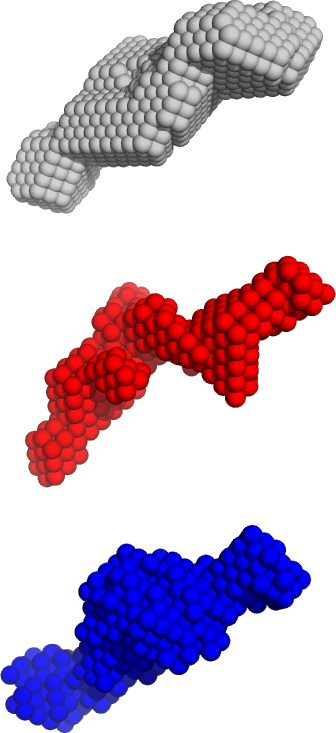}
  \caption{Point cloud registration onto SASDAA5 (grey top). Bead models SASDBS6 (red) and SASDBY6 (blue) were fitted onto SASDAA5 by maximizing the kernel correlation using DAMM.}\label{fig:SASDAA5}
\end{figure}

The second example involves fitting a bead model of Human Chromatin Remodeler CHD4 (SASBDB code SASDAA5) against two other SASBDB structures: the apo form of full length ObgE (SASDBS6) and mitochondrial heat shock protein 70 (SASDBY6). The superposition is shown in figure \ref{fig:SASDAA5}.

\clearpage

\section{Kernel correlation as a proxy for the cross-correlation coefficient and the RMSD}
By construction, the kernel correlation is highly related to the cross-correlation coefficient (CCC) that is often used to measure the overlap of two cryo-EM density maps represented on voxel grids. This is demonstrated in the figure below for one of the cryo-EM docking targets discussed in the main text (superposition of two RNA polymerase II structures). We observe a high correlation between the negative log kernel correlation (which is optimized by the MM algorithms introduced in Methods) and CCC between the target map and the transformed map. Therefore, by minimizing $-\log\KC$ we effectively maximize the CCC between the two density maps. 

A high correlation is also observed between the RMSD (as defined in equation 7 of the main article) and the negative log kernel correlation. The RMSD is optimized by ICP. 

\begin{figure}[hbt!]
  \centering
  \includegraphics[width=0.85\textwidth]{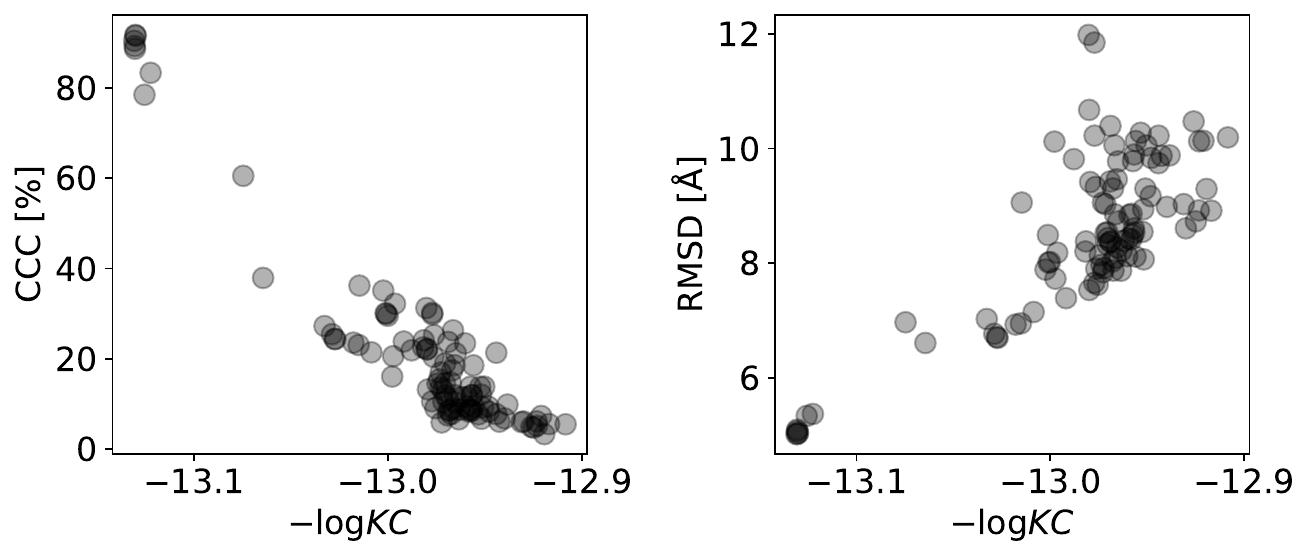}
  \caption{Kernel correlation as a proxy for RMSD and cross-correlation coefficient (CCC). Left: High correlation between $-\log{} \KC$ and CCC. Right: High correlation between  $-\log{} \KC$ and the RMSD (see equation 7 in main manuscript.)}
\end{figure}

\clearpage

\section{Docking the high-resolution structure of a subunit into a cryo-EM map of a symmetric assembly}
The first point cloud, $(\vec{X}, \vec{q})$, represents the full assembly and is derived, for example, from a cryo-EM map. The second point cloud, $(\vec{Y}, \vec{p})$, represents the subunit that will be docked into the full assembly. We assume that the assembly is symmetric and the symmetry mates can be generated from a single subunit by the action of $C$ rigid transformations $\{(\R_k, \T_k)\}_{k=1}^C$. The subunit needs to be transformed by an unknown rigid transformation $(\R, \T)$ such that the overlap between the target and the full model structure is as large as possible. The coordinates of the $k$-th subunit after rigid transformation are:
\[
\Y_{jk}' = \R_k (\R \Y_j + \T) + \T_k
\]
The kernel correlation between the point cloud representing the assembly and the structure of the assembly built by applying the symmetry operators is:
\[
\KC_{\text{sym}}(\R, \T) = \sum_{i=1}^M \sum_{j=1}^N \sum_{k=1}^C q_i p_j \phi_\sigma(\|\X_i - \R_k (\R \Y_j + \T) - \T_k\|)
\]
We can rewrite the kernel correlation of a symmetric assembly:
\[
\KC_{\text{sym}}(\R, \T) = \sum_{i=1}^M \sum_{j=1}^N \sum_{k=1}^C q_i p_j \phi_\sigma(\|\R_k^T(\X_i - \T_k) - \R \Y_j - \T\|)
\]
Upper bound:
\begin{eqnarray*}
  - \log \KC_{\text{sym}}(\R, \T)
  &=&
  - \log \left\{\sum_{i=1}^M \sum_{j=1}^N \sum_{k=1}^C q_i p_j \phi_\sigma(\|\X_i - \R_k (\R \Y_j + \T) - \T_k\|)\right\} \\
  &=&
  - \log \left\{\sum_{ijk} q_i p_j w_{ijk} \frac{\phi_\sigma(\|\X_i - \R_k (\R \Y_j + \T) - \T_k\|)}{w_{ijk}} \right\} \\
  &\le&
  \frac{1}{2\sigma^2} \sum_{ijk} q_i p_j w_{ijk} \, \|\X_i - \R_k (\R \Y_j + \T) - \T_k\|^2 + \mathrm{const.} \\
\end{eqnarray*}
where the weights
\[
w_{ijk} \propto \phi_\sigma(\|\X_i - \R_k (\R \Y_j + \T) - \T_k\|)
\]
are normalized such that $\sum_{ijk} q_i p_j w_{ijk} = 1$.

More examples of 3D fitting into symmetric assemblies by maximizing $\KC_{\text{sym}}$ are shown in Fig. \ref{fig:symmetric}.

\begin{figure}[hbt!]
  \centering
  \subfigure[\centering EMD-6422]{
    \includegraphics[width=0.9\textwidth]{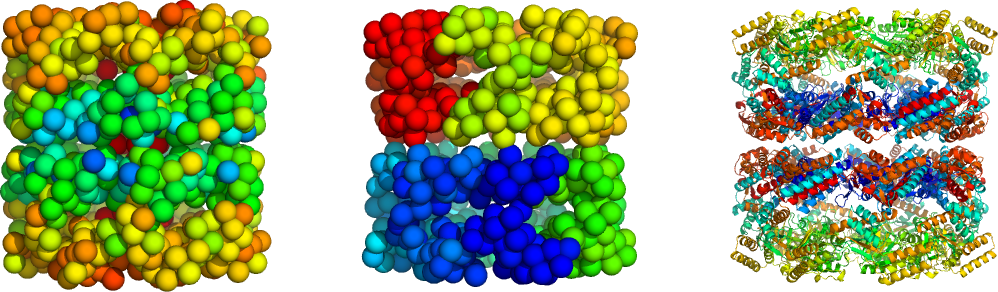}}
  \qquad
  \subfigure[\centering EMD-5995]{
    \includegraphics[width=0.9\textwidth]{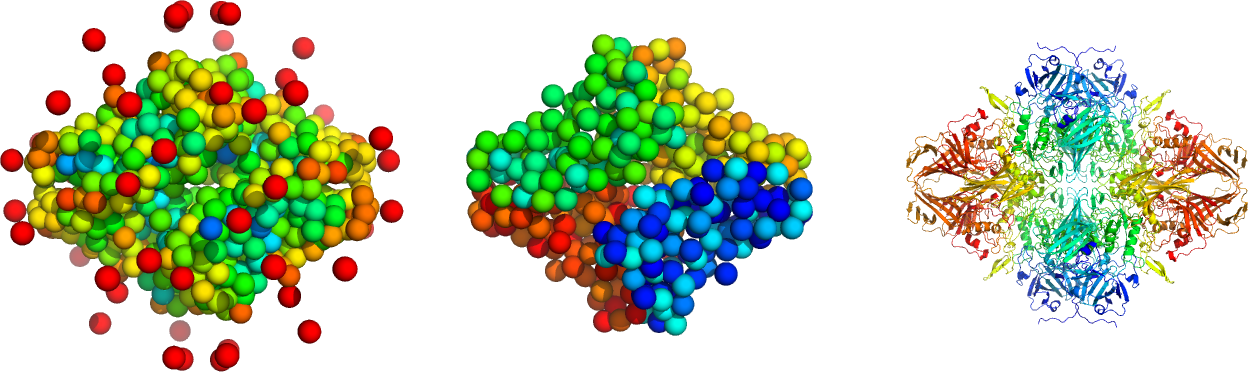}}
  \qquad
  \subfigure[\centering EMD-6000]{
    \includegraphics[width=0.9\textwidth]{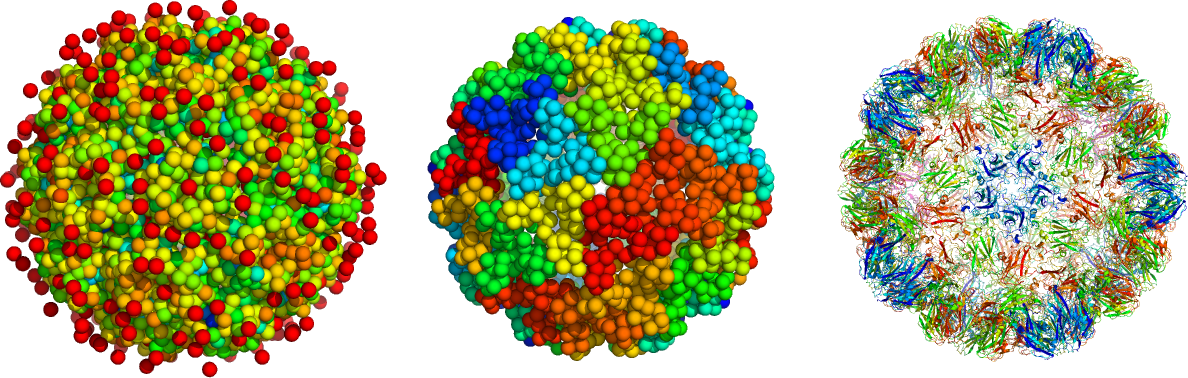}}
  \caption{Rigid docking of a subunit into a symmetric assembly. EMD-6422: GroEL, D7 symmetry. EMD-5995: beta-galactosidase, D2 symmetry. EMD-6000: Brome mosaic virus. In each row, the point cloud on the left shows the particle representation of the cryo-EM map of the assembly obtained by running DP-means with a particle radius of 5 \AA. The colors indicate the weight of the particles. The middle panel shows the fitted structure of the subunit that was also coarse-grained by running DP-means. The structure in blue is the subunit; all other particles were generated on the fly by applying the symmetry operators. The right panel is the high-resolution structure of the full assembly obtained by transforming the high-resolution structure of the subunit rigidly by using the estimated pose and by generating the symmetry mates by applying the symmetry operators.}\label{fig:symmetric}
\end{figure}

\end{appendix}

\end{document}